\documentclass[a4paper,10pt,aps,pre,twocolumn,amsmath,amssymb,nofootinbib]{revtex4}
\usepackage{graphicx,soul}
\usepackage{multirow}
\usepackage[dvipsnames]{xcolor}
\usepackage[colorlinks=true,linkcolor=blue,citecolor=red]{hyperref}%

\newcommand{\dd}{\mathrm{d}}

\newcommand{\pd}[2]{\frac{\partial #1}{\partial #2}}

\newcommand{\mean}[1]{\langle #1 \rangle}
\newcommand{\Int}[1]{\int\dd #1\;}
\newcommand{\IInt}[3]{\int_{#2}^{#3}\dd #1\;}
\newcommand{\x}{\mathbf r}
\newcommand{\X}{\mathbf R}
\newcommand{\vhi}{\varphi}

\newcommand{\RNum}[1]{\uppercase\expandafter{\romannumeral #1\relax}}

\DeclareMathOperator{\erf}{erf}

\begin{document}

\title{Crystallization of hard spheres revisited. \RNum{2}. Thermodynamic modeling, nucleation work, and the surface of tension}

\author{David Richard}
\affiliation{Institut f\"ur Physik, Johannes Gutenberg-Universit\"at Mainz, Staudingerweg 7-9, 55128 Mainz, Germany}
\author{Thomas Speck}
\affiliation{Institut f\"ur Physik, Johannes Gutenberg-Universit\"at Mainz, Staudingerweg 7-9, 55128 Mainz, Germany}

\begin{abstract}
  Combining three numerical methods (forward flux sampling, seeding of droplets, and finite size droplets), we probe the crystallization of hard spheres over the full range from close to coexistence to the spinodal regime. We show that all three methods allow to sample different regimes and agree perfectly in the ranges where they overlap. By combining the nucleation work calculated from forward flux sampling of small droplets and the nucleation theorem, we show how to compute the nucleation work spanning three orders of magnitude. Using a variation of the nucleation theorem, we show how to extract the pressure difference between the solid droplet and ambient liquid. Moreover, combining the nucleation work with the pressure difference allows us to calculate the interfacial tension of small droplets. Our results demonstrate that employing bulk quantities yields inaccurate results for the nucleation rate.
\end{abstract}

\maketitle

\section{Introduction}

The accurate and reliable prediction of nucleation rates and pathways is still a major challenge in computational chemistry. Classical nucleation theory (CNT)~\cite{skripov1974metastable,kashchiev2000nucleation} provides a simple thermodynamic framework to compute the nucleation rate based on the reversible nucleation work required to form a droplet of the thermodynamically stable phase surrounded by the metastable phase. However, standard CNT is based on two crucial assumptions: (i)~it employs bulk values for the pressure difference between the two phases, and (ii)~in CNT one assumes the capillary approximation, where the interfacial tension equals the bulk value $\gamma_\infty$ of a flat interface. In practice, the interfacial tension is often treated as a free fit parameter yielding values that might significantly divert from the corresponding bulk values.

In part~\RNum{1} we have revisited the crystallization of (almost) hard spheres and presented a method to extract the free energy landscape from dynamically biased forward flux simulations (FFS). We have reproduced nucleation rates from previous simulations~\cite{auer2004numerical,filion2010crystal,dijkstra11}, which yield an effective interfacial tension $\gamma_\text{CNT}\simeq0.76$~\cite{espinosa2016seeding,speckpartone}. This value is 35\% too large compared to $\gamma_\infty\simeq0.56$ (in units of thermal energy and particle diameter). As reference for the interfacial tension, we use the results obtained in Ref.~\cite{schmitz2015ensemble}.

The purpose of this second part is to revisit the thermodynamic modelling underlying CNT and to investigate so-called ``non-classical effects''. Non-classical effects that are not included in CNT are: (i) The change of density at the center of the droplet~\cite{oxtoby1988nonclassical} and the loss of bulk properties. These observations were confirmed in simulation for gas-liquid~\cite{ten1998computer} and solid-liquid~\cite{russo2012microscopic} nucleation. (ii) The variation of the interfacial tension due to the curvature~\cite{tolman1949effect,joswiak2016energetic}, which includes an energetic contribution since the number of bonds of a curved interface differs from a flat interface~\cite{joswiak2016energetic} and an entropic contribution due to fluctuations of the interface (capillary waves)~\cite{vink2005capillary,schmitz2014logarithmic}. (iii) Additional entropic effects such as translational entropy and domain breathing \cite{schmitz2014logarithmic}. These effects were recently highlighted to be crucial for computing the interfacial tension of flat interfaces~\cite{schmitz2014determination,benjamin2015crystal}. The extension to curved interfaces is discussed in Ref.~\cite{troester2017equilibrium}. Finally, (iv) the shape of the droplet~\cite{jungblut2011crystallization,prestipino2013fingerprint} and the anisotropy of the interfacial tension when one deals with solid, faceted nuclei~\cite{ringe2011wulff}.

Here we turn to computer simulations of a model with short-range continuous forces that can be mapped onto true hard spheres (cf. part~\RNum{1}). During the last 20 years, computer simulations have been shown to be well-suited to explore non-classical~\cite{nijmeijer1992molecular,ten1998computer,macdowell2004evaporation,troster2011positive,troster2012numerical,macdowell2013capillary,schmitz2014determination,schmitz2014logarithmic,benet2015interfacial,hofling2015enhanced,espinosa2016seeding,troester2017equilibrium} and other features~\cite{van2009hard,russo2013interplay,ni2014crystallizing,richard2015role}. Unbiased simulations are confronted with the fact that nucleation is a rare event, and only rather large supersaturations can be studied, for which the spontaneous formation of critical droplets occurs within an acceptable time. To close the gap to coexistence, we will employ three different methods that allow to study larger and larger droplets. The first one is based on preparing an initial metastable state and evaluating the nucleation barrier through rare event sampling, which has been the focus of the first part of this series. This is done either employing constrained~\cite{quigley2008metadynamics,filion2010crystal} or, as presented in part~\RNum{1}, unconstrained dynamics~\cite{van2005elaborating,wedekind2008kinetic,valeriani2007computing}. These simulations lead to the determination of the nucleation work $\Delta F_c$, which can then be used to quantify possible deviations from the classical picture on CNT. The second approach is based on the ``seeding'' of droplets~\cite{sanz2013homogeneous,espinosa2014homogeneous,zimmermann2015nucleation,espinosa2016seeding,lifanov2016nucleation}, whereby an initial droplet is placed into the metastable liquid at given supersaturation. This initial seed either grows or dissolves, with the crossing determining the size of the critical droplet $n_c$. The seeding method allows to study larger droplets than accessible in FFS, but does not allow to evaluate the nucleation rate nor the nucleation work directly. To gain access to the work, one approach has been to reverse the CNT expression for $n_c$ to evaluate the corresponding effective interfacial tension $\gamma_\text{CNT}$~\cite{espinosa2016seeding}. Finally, the third method is to exploit the coexistence of a droplet in equilibrium with the surrounding liquid~\cite{binder1987theory,macdowell2004evaporation,troster2011positive,troster2012numerical,statt2015finite}. After a transient time the time evolution of the droplet size reaches an average value around $n_c$, and further growth is inhibited by the finite size of the simulation box. Also with this method the nucleation work is not accessible directly.

In the following, we exploit the nucleation theorem to calculate the nucleation work $\Delta F_c$ and pressure difference for a wide range of droplet sizes down to very small supersaturations. This is achieved through extracting the excess number of particles, which is an unambiguous thermodynamic quantity and easily accessible in all described simulation methods and does not require any assumptions. Moreover, knowing exactly the nucleation work and size-dependent pressure allows to evaluate the corresponding interfacial tension of droplets. We demonstrate that the behavior of droplets sampled with different methods overlap and agree with each other. Both show significant deviations for finite droplets from the respective bulk values.

\section{Methods}

\subsection{Simulations}

We perform molecular dynamics (MD) simulations of a one-component system interacting through the repulsive Weeks-Chandler-Andersen potential, which is mapped onto hard spheres at packing fraction $\phi$ through an effective diameter. Details of this mapping and the forward flux simulations (FFS) can be found in part~\RNum{1}~\cite{speckpartone}. From sampled configurations, we identify particles as solid employing a bond-order parameter (described in appendix~\ref{ap:bop}) and construct clusters of mutually bonded solid particles. We focus on the largest cluster with size $n$ and discard information on the other clusters. It is important to bear in mind the difference of the cluster size $n$, which is an order parameter, from the thermodynamic variables $N_i$ to be introduced in the next Sec.~\ref{sec:therm}.

In addition to forward flux sampling, we perform simulations in which solid droplets are ``seeded'' and simulations of (metastable) finite-size droplets. For these simulations we use the LAMMPS package with underdamped Langevin dynamics in the NVT ensemble~\cite{plimpton1995fast}. Throughout this work, we employ dimensionless quantities with lengths given in units of the effective diameter and energies (such as the chemical potential) in units of the thermal energy $k_\text{B}T$.

\subsection{Thermodynamics of nucleation}
\label{sec:therm}

\subsubsection{Dividing surface}

The thermodynamic treatment of inhomogeneous systems has a long history and goes back to Gibbs~\cite{gibbs1948collected}. Here, we consider the transformation of an initially homogeneous liquid to an inhomogeneous system comprising a solid droplet in the surrounding melt. The system is composed of $N$ particles, and the transformation is assumed to occur at constant volume $V$ and constant temperature $T$. Pivotal is the concept of a \emph{dividing surface} (DS), which divides the system into two disjunct volumes with constant densities: the volume occupied by the droplet $V_s$ and the remaining volume for the liquid phase $V_l=V-V_s$. Formally, we distinguish three particle species: $N_s$ solid particles, $N_l$ liquid particles, and $N_x$ particles that are attributed (adsorbed) to the interface. The conservation of the number of particles imposes $N=N_s+N_l+N_x$. We stress that the dividing surface is not unique and several prescriptions exist \cite{kashchiev2003thermodynamically}.

\begin{figure}[t]
  \includegraphics[scale=1.0]{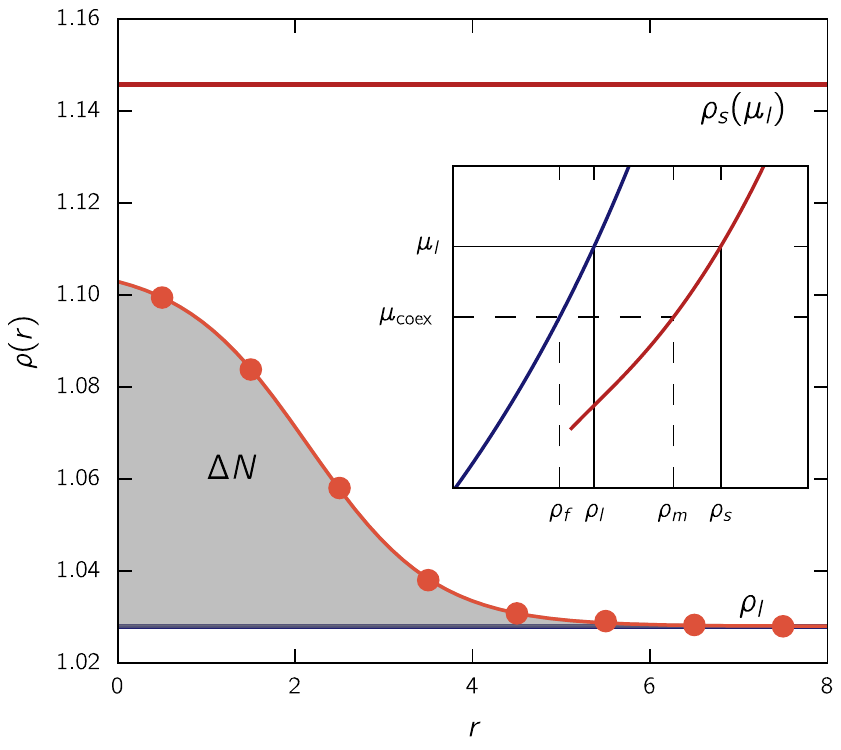}
  \caption{\textbf{Density profile from simulations.} Average radial density distribution $\rho(r)$ (symbols and line) with respect to the center of mass of droplets sampled from FFS close to the transition state at packing fraction $\phi\simeq0.539$. The blue and red horizontal lines indicate the bulk liquid and solid densities at the same chemical potential $\mu_l$ of the metastable liquid. The gray area shows the excess number of particle $\Delta N$. The inset shows the bulk equations of state for liquid (blue) and solid (red). At the same ambient chemical potential $\mu_l$, it defines the bulk densities $\rho_l$ and $\rho_s$. At coexistence, these are referred to as freezing and melting density (dashed lines).}
  \label{fig:nucleus}
\end{figure}

Fig.~\ref{fig:nucleus} shows a radial density profile $\rho(r)$ from forward flux simulations averaged over droplets with the same number $n$ of solid particles, for details see appendix~\ref{ap:profile}. In contrast to the idealized concept of a sharp dividing surface, we observe a gradual change from the center of the droplet at $r=0$ to the liquid phase with density $\rho(r\to\infty)=\rho_l$. While this density agrees with the bulk value, the density $\rho(0)$ at the center is substantially smaller than the corresponding bulk density $\rho_s$ of a solid at the same ambient chemical potential $\mu_l$ (this was already pointed out by Gibbs~\cite{gibbs1948collected}). Clearly, thermodynamic quantities like the interfacial tension will depend on the specific way we replace this actual density profile by a DS.

At constant temperature, the reversible work $W=\Delta F$ for the formation of such a droplet is the free energy difference
\begin{equation}
  \label{eq:fe}
  \Delta F = (\mu_s-\mu_l)N_s +(\mu_x-\mu_l)N_x -(P_s-P_l)V_s + \Phi
\end{equation}
between the inhomogeneous system and the homogeneous liquid with $F_0=\mu_lN-P_lV$. Here, $\mu_i$ (with $i=s,l,x$) are the chemical potentials, $P_s$ and $P_l$ are the pressures associated with the solid droplet and the liquid phase, respectively, and $\Phi$ is the excess free energy due to the interface. By construction, the DS has no volume and, consequently, no corresponding pressure term. In writing down Eq.~(\ref{eq:fe}) we made the assumption that the chemical potential of the homogeneous liquid is the same as that of the liquid in the inhomogeneous system.

Geometric scaling dictates that the dependence of the free energy difference on the droplet volume $V_s$ is non-monotonous, it increases for small volumes due to the cost of the interface and decreases at large volumes as the lower bulk free energy of the solid phase dominates. The maximum of the free energy barrier defines an ensemble of transition states containing critical droplets. Albeit unstable, these droplets are stationary and thus there is no particle flux between the liquid and the solid droplet, which implies that the corresponding chemical potentials are equal, $\mu_l=\mu_s=\mu_x$~\cite{guggenheim1940thermodynamics}. Irrespective of the definition of the DS, at the top of the barrier the derivative $\partial \Delta F / \partial V_s=0$ of Eq.~(\ref{eq:fe}) vanishes and we find
\begin{equation}
  \label{eq:lap}
  \Delta P = P_s - P_l = \frac{\partial\Phi}{\partial V_s},
\end{equation}
which relates the pressure difference driving the phase transformation to the excess surface free energy.

We now fix the shape of critical droplets to be spherical with radius $R$.  Even if the actual structure of solid droplets, \emph{e.g.} defined from bond order parameters, would be described by an object resembling a faceted crystal, one can still unambiguously define a radial density profile as shown in Fig.~\ref{fig:nucleus}, and treat $R$ as an effective parameter. Moreover, we assume that the ensemble of critical droplets is fully described by two quantities, the radius $R$ and the number of particles $N_x$. Hence, we can write the excess free energy as $\Phi_c=\Phi_c(R,N_x)$ which we split into
\begin{equation}
  \label{eq:phi}
  \Phi_c(R,N_x) = \gamma(R,N_x)A(R)
\end{equation}
with interfacial tension $\gamma$ and area $A=4\pi R^2$ of the dividing surface. Eq.~(\ref{eq:lap}) then becomes the (generalized) Laplace equation
\begin{equation}
  \label{eq:lap:sphere}
  \Delta P = \frac{2\gamma}{R}+\pd{\gamma}{R}
\end{equation}
valid for any spherical DS. As typically done in the context of nucleation, we will consider two choices for the DS: (i)~The \emph{surface of tension}~\cite{tolman1948consideration,rowlinson2013molecular} with radius $R_s$ defined through $\partial\gamma/\partial R|_{R_s}=0$ at which the interfacial tension $\gamma_s=\gamma(R_s)$ is minimal, and (ii)~the equimolar dividing surface (EDS) with radius $R_e$ defined through $N_x=0$. The difference $R_e-R_s$ between these radii is related to the Tolman length~\cite{tolman1948consideration}.

\subsubsection{Classical nucleation theory}

We now seek an expression for the nucleation work $\Delta F_c$ to form a critical droplet. From Eq.~(\ref{eq:fe}) we find
\begin{equation}
  \label{eq:w:cnt}
  \Delta F_c = -\Delta PV_s + \Phi_c = \frac{\Delta PV_s}{2}
  = \frac{16\pi\gamma_s^3}{3(\Delta P)^2}
\end{equation}
eliminating area and volume with the help of the Laplace equation~(\ref{eq:lap:sphere}) employing the surface of tension. This result agrees with the expression for the nucleation work employed in CNT, which thus fixes the DS of CNT to the surface of tension. Note that Eq.~(\ref{eq:w:cnt}) is an exact result, but it is not very useful to predict the nucleation work since we do not know the interfacial tension $\gamma_s$ and pressure difference $\Delta P$. The most common approximation is to assume the interfacial tension to correspond to that of a flat interface $\gamma_\infty=\gamma(R\to\infty)$ (the capillary approximation), and to employ the pressure difference $\Delta P_\infty=P_s(\rho_s)-P_l(\rho_l)$ between a droplet at solid bulk density $\rho_s$ and the surrounding liquid. Clearly, both approximations are questionable and one should not be surprised to observe deviations of the actual nucleation work from the thus predicted $\Delta F_{c,\infty}$. However, this is not a failure of classical nucleation theory nor the thermodynamic modeling but due to the uncontrolled and inappropriate approximation of interfacial tension and pressure difference.

One often encounters a form of the nucleation work that differs from Eq.~(\ref{eq:w:cnt}). It is obtain through integrating the solid branch $\partial\mu_s/\partial P_s=V_s/N_s$ assuming the solid density $N_s/V_s$ to be independent of pressure. This leads to $\mu_s(P_s)-\mu_s(P_l)=\Delta P V_s/N_s$ and using $\mu_s(P_s)=\mu_l(P_l)$ to
\begin{equation}
  \label{eq:w:mu}
  \Delta F_c = \frac{[\mu_l(P_l)-\mu_s(P_l)]N_s}{2}
\end{equation}
expressing the nucleation work as the difference of chemical potential between liquid and solid at the ambient pressure $P_l$.

\subsubsection{Nucleation theorem}

There is an alternative way to calculate the nucleation work from the excess number of particles
\begin{equation}
  \Delta N = 4\pi\int_0^{\infty}dr \;r^2[\rho(r)-\rho_l].
  \label{eq:nexcess}
\end{equation}
Density profiles of droplets can be measured very precisely in computer simulations as we demonstrate below. They have a well-defined meaning independent of the choice of bond-order parameter and the dividing surface. The connection between the critical excess $\Delta N_c$ and the reversible work $\Delta F_c$ associated with creating a critical droplet was first derived from thermodynamic arguments by Kaschiev \cite{kashchiev1982relation,oxtoby1994general} and confirmed later by Bowles \emph{et al.}~\cite{bowles2000molecular}.

We again start from $\Delta F_c=-\Delta PV_s+\Phi_c$ but now take the derivative with respect to the chemical potential $\mu_l$,
\begin{equation}
  \label{eq:fe:mu}
  \pd{\Delta F_c}{\mu_l} = -\pd{\Delta P}{\mu_l}V_s + \pd{\Phi_c}{\mu_l} 
  + \left(-\Delta P+\pd{\Phi_c}{V_s}\right)\pd{V_s}{\mu_l},
\end{equation}
where we use $V_s=V_s(\mu_l)$. With Eq.~(\ref{eq:lap}), the last term cancels. Note that formally we have changed the ensemble [in particular $\Phi=\Phi(\mu_x)$] and now control the chemical potentials with fluctuating particle number in order to be able to describe the droplet as a density fluctuation. In simulations the total number of particles is conserved and, therefore, the formation of a droplet leads to a (slight) reduction of $\rho_l$.

Employing the partial derivatives
\begin{equation}
  \pd{\Phi_c}{\mu_x} = -N_x, \qquad \pd{P_i}{\mu_i} = \frac{N_i}{V_i}
\end{equation}
with $i=s,l$, and exploiting that the chemical potentials are equal for critical droplets, we eliminate the pressure to obtain the nucleation theorem
\begin{equation}
  \label{eq:theorem}
  \pd{\Delta F_c}{\mu_l} = -\left(N_s+N_x-N_l\frac{V-V_l}{V_l}\right) 
  = -\Delta N_c.
\end{equation}
Here, $\Delta N_c=N_s+N_l+N_x-\rho_lV$ is the number of additional particles present for critical droplets compared to a homogeneous liquid with density $\rho_l=N_l/V_l$ at volume $V$. This relation can be used to evaluate $\Delta F_c$ for a large range of supersaturations by computing the function $\Delta N(\mu_l)$ and evaluating, for one reference state point $\mu_0$, the nucleation work $\Delta F_c(\mu_0)$ (for example using FFS)~\cite{kashchiev2000nucleation},
\begin{equation}
  \Delta F_c(\mu_l) = \Delta F_c(\mu_0) - \IInt{\mu}{\mu_0}{\mu_l}\Delta N_c(\mu).
  \label{eq:ntheorem}
\end{equation}
DFT calculations have shown that Eq.~(\ref{eq:theorem}) holds even for very small droplets~\cite{oxtoby1994general,hruby2007gradient}.

We are still missing an expression that allows to evaluate the thermodynamic pressure associated with the droplet. To this end, we equate Eq.~(\ref{eq:fe:mu}) with Eq.~(\ref{eq:theorem}) to obtain
\begin{equation}
  \pd{\Delta P}{\mu_l} = \frac{\Delta N_c-N_x}{V_s}.
\end{equation}
Integrating this expression along the equimolar radius $R_e$ we have $N_x=0$ by definition and thus the pressure difference becomes
\begin{equation}
  \label{eq:pressure}
  \Delta P(\mu_l) = \IInt{\mu}{\mu_\text{coex}}{\mu_l}
  \frac{\Delta N_c(\mu)}{V_s(R_e(\mu))} 
  = \IInt{\mu}{\mu_\text{coex}}{\mu_l} \Delta\rho(\mu)
\end{equation}
since $\Delta P$ vanishes at coexistence, where
\begin{equation}
  \label{eq:ratio}
  \Delta\rho(\mu) = \frac{\Delta N_c(\mu)}{V_s(R_e(\mu))}
  = \rho(0;\mu)-\rho_l(\mu).
\end{equation}
Eqs.~(\ref{eq:ntheorem},\ref{eq:pressure}) allow to calculate both the nucleation work and the pressure difference from density profiles of droplets sampled at different global packing fractions. Rearranging the exact expression Eq.~(\ref{eq:w:cnt}) for the nucleation work, we can thus extract the (thermodynamic) interfacial tension $\gamma_s$ of critical droplets as a function of $\mu_l$ or, equivalently, packing fraction $\phi$.

\subsection{Seeding of droplets}
\label{sec:seeding}

\begin{figure}[b!]
  \includegraphics[scale=.97]{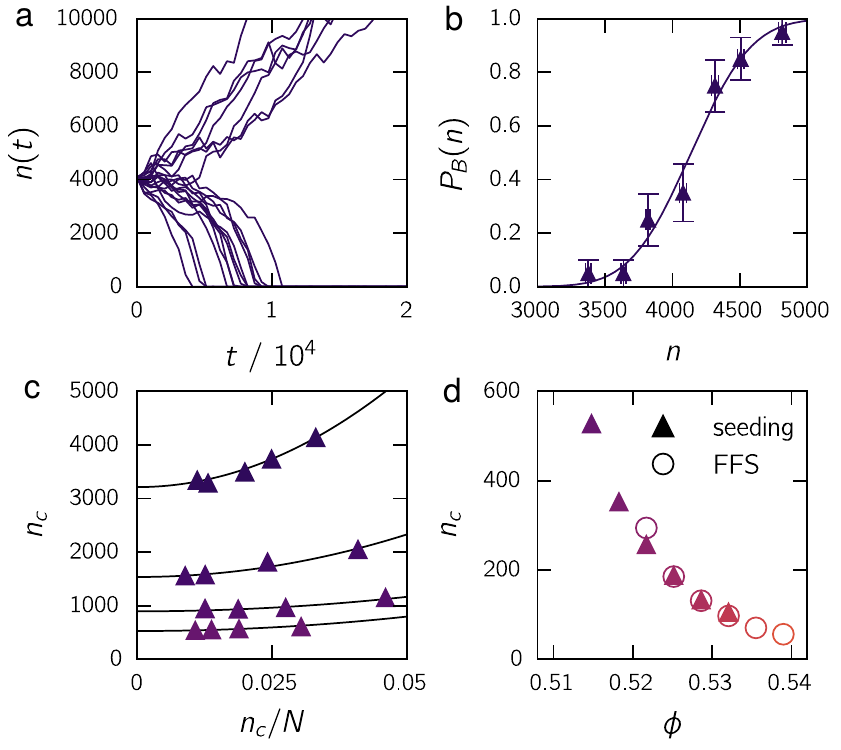}
  \caption{\textbf{Seeding of droplets.} (a)~Evolution of the droplet size for 20 independent fleeting trajectories starting from a seed with $n\simeq 4,000$. (b)~Committor probability $P_B$ as a function of the seed size $n$. The solid line indicates the fit to determine $n_c$ at $P_B=1/2$. (c)~Critical droplet size $n_c$ as a function of the ratio $n_c/N$. Solid lines are fits with $n_c(N)=n_c(\infty)+a(n_c/N)^2$ to extract the limiting critical size $n_c(\infty)$. (d)~Dependence of the critical size as a function of the packing fraction $\phi$ for medium and small droplets. The filled triangles and empty circles are the data from the seeding method and FFS, respectively.}
  \label{fig:seeding}
\end{figure}

Seeding methods aim to probe saturations close to coexistence, which are not accessible by rare event sampling methods~\cite{sanz2013homogeneous,zimmermann2015nucleation,espinosa2016seeding}. Here, a variant of the original proposed seeding method from Ref.~\cite{espinosa2016seeding} is employed to extract the critical droplet size of a supersaturated liquid at pressure $P_l>P_{\text{coex}}$. For each packing fraction $\phi$, a spherical crystal seed is placed into a surrounding melt. The density of the initial solid droplet is taken to have the same chemical potential $\mu_l(\phi)$ as the liquid. We then use the Clarke-Wiley algorithm to remove all particle overlaps \cite{Wiley87}. After this procedure, the droplet size is $n_0$ and we start short warmup runs ($t_{wp}=2.5-5$) from this configuration to reach the equilibrium pressure $P_l(\phi)$ of the metastable liquid and measure the droplet size $n$. We then run 20 independent fleeting trajectories to determine whether the droplet grows or melts. The typical time evolution of the cluster size $n(t)$ is shown in Fig.~\ref{fig:seeding}(a).

We compute the probability $P_A(n_0)$ to return to basin $A$ (the melt without a droplet) as the fraction of trajectories that reach the threshold $n<10$. The committor probability follows as $P_B(n)=1-P_A(n)$ and is fitted with the function $P_B(n) = \frac{1}{2}[1+\erf(a(n-n_c))]$ to extract the critical droplet size $n_c$ at which $P_B=1/2$, more details are provided in part~\RNum{1}. In Fig.~\ref{fig:seeding}(b), we present this procedure for $\phi\simeq0.504$ and $N=125,000$. Since our simulations are perform in NVT, in a finite system there is a depletion of liquid particle and thus a reduction of the liquid density and the pressure imposed on the droplet, which results in larger critical droplets. To overcome this issue and extract the thermodynamic limit for $n_c$ at a given $\phi$, we run simulations for different system sizes and fit $n_c(N)=n_c(\infty)+a(n_c/N)^2$ with free parameter $a$, see Fig.~\ref{fig:seeding}(c). Moreover, we check for smaller critical droplet sizes that the seeding method matches our previous results obtained from FFS. In Fig.~\ref{fig:seeding}(d), we show the evolution of $n_c$ as a function of the packing fraction $\phi$. We find a perfect agreement between the seeding method and FFS, with one small deviation for the lower packing fraction extracted in FFS ($\phi\simeq 0.52$). This discrepancy is due to finite size effects for the FFS data, where the system is composed of $N=5,000$ particles. This deviation will be corrected in Sec.~\ref{sec:result} by computing the chemical potential of the coexisting liquid to extract the true supersaturation $\Delta\mu=\mu_l-\mu_\text{coex}$ of the metastable liquid.

\subsection{Finite-size droplets}
\label{sec:fsd}

Although the seeding method can probe large critical droplet sizes, it suffers from the need to increase the system size as the supersaturation is decreased in order to avoid finite-size effects. Moreover, for large seeds one has to increase the fleeting time to let the system melt if it is sub-critical. An alternative to avoid these performance issues is to directly look at equilibrium droplets, which are present in a finite system at coexistence~\cite{binder1987theory,macdowell2004evaporation,isobe2015hard,koss2017free}. It is well-known that when a finite system is compressed (or cooled) above its freezing point $\phi_f$, the homogeneous liquid has still a lower free energy than any phase separated state. This stability continues until the droplet transition ($\phi_t>\phi_f$), above which a solid droplet in coexistence with a liquid melt becomes more stable. In this new state, the two phases have the same chemical potential $\mu_l=\mu_s$ (chemical equilibrium), but the pressure inside the solid droplet exceed the one in the liquid phase, $P_s>P_l$, to compensate the presence of an interfacial tension (mechanical equilibrium). Employing periodic boundary conditions, this transition is then followed by the droplet-to-cylinder and cylinder-to-slab transitions, which geometrically minimize the surface free energy cost. For the slab, the two phases have both the same pressure and chemical potential, and the two coexisting packing fractions are $\phi_f$ and $\phi_m$, respectively. While every transition is interesting in its own~\cite{macdowell2004evaporation,moritz2017interplay}, we will focus in this paper on the droplet transition. The scaling at which this transition occurs is controlled by the system dimension and size, and follow in 3 dimensions \cite{binder1987theory},
\begin{equation}
\phi_t-\phi_f\sim V^{-1/4}.
\end{equation}
We emphasis that at $\phi>\phi_t$ the solid droplet is more stable than the homogeneous state, but it can still be metastable with respect to the cylinder and slab.

One can, without performing any simulations, produce the stability range of the different geometries by assuming the capillary approximation for the interfacial tension, $\gamma(V_s)\approx\gamma_{\infty}$ and the bulk equations of state. To this end, we minimize the free energy difference 
\begin{multline}
  \Delta F(\rho,\rho_l,\rho_s) 
  = [\mu_l(\rho_l)-\mu_l(\rho)]N-[P_l(\rho_l)-P_l(\rho)]V \\
  + ([P_s(\rho_s)-P_l(\rho_l)]V_s + \gamma_{\infty}\alpha V_s^n
  \label{eq:diff}
\end{multline}
between the homogeneous state at chemical potential $\mu_l(\phi)$ and pressure $P_l(\phi)$ in coexistence with a solid domain at the same chemical potential $\mu_l(\phi_l)$ but different pressure $P_s$. Here, the prefactor $\alpha$ and the exponent $n$ describe the shape of the solid domain with
\begin{align}
  \text{sphere: } &\alpha = (36\pi)^{1/3},\ n=2/3 \\
  \text{cylinder: } &\alpha = 2\sqrt{\pi L},\ n=1/2 \\
  \text{slab: } &\alpha = L^2,\ n=0.
\end{align}
Note that in Eq.~(\ref{eq:diff}) we use the equimolar dividing surface, which sets $N_x=0$. We can directly evaluate for the EDS the solid volume $V_s$ via the lever rule $\phi V=\phi_l(V-V_s)+\phi_sV_s$. The first step of this minimization is to find for each shape which liquid density $\rho_l$ minimizes $\Delta F$. The second step is to find the global minimum of $\Delta F$ by comparing the free energy of the homogeneous liquid with the free energy of the three different geometries.

\begin{figure}[t]
\includegraphics[scale=1.0]{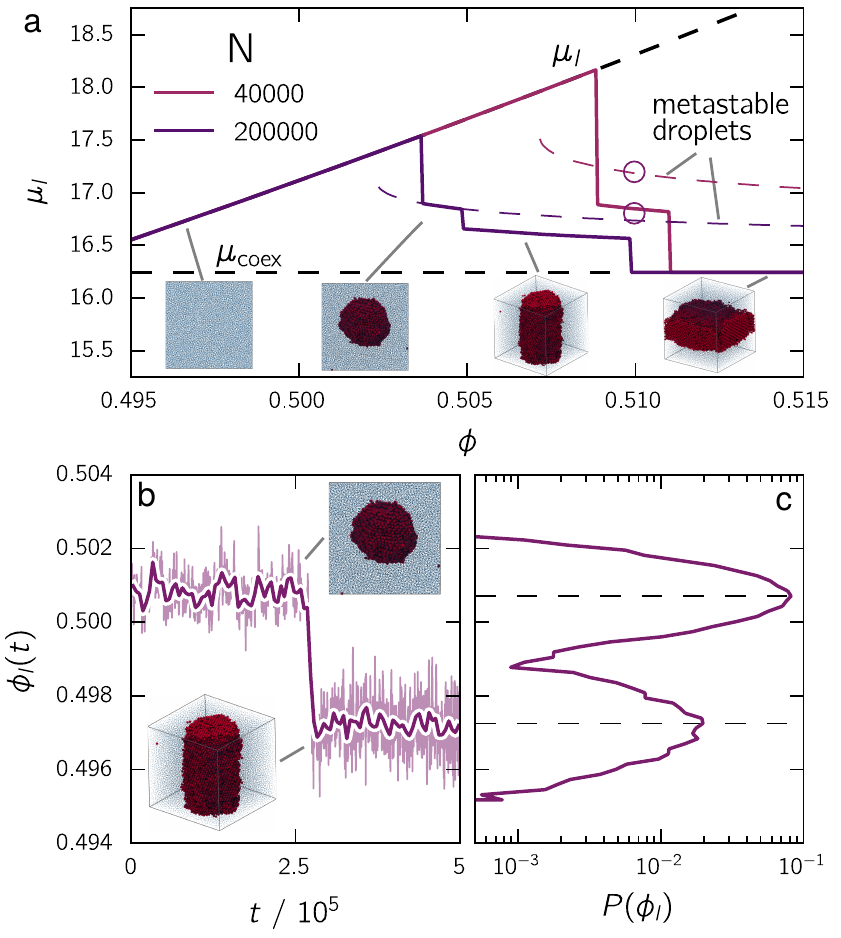}
\caption{\textbf{Finite-size droplets.} (a)~Theoretical prediction for the chemical potential $\mu_l$ of the liquid as a function of the global packing fraction $\phi$ in the capillary approximation. The solid lines indicate the thermal equilibrium, and the dashed lines the metastable droplet branches. The colors distinguish two different system sizes, namely $N=4\times10^4$ and $N=2\times10^5$. Snapshots illustrate the different geometries of the solid phase. (b)~Time evolution of the packing fraction of the liquid phase $\phi_l(t)$ during a transition from a metastable droplet to an equilibrium cylinder phase ($N=4\times10^4$). Snapshots indicate the two different geometries. (c)~Average probability distribution of $\phi_l$ for 10 independent runs. The two dashed lines indicate the average liquid packing fraction in the droplet and cylinder phase. We convert these two densities into chemical potentials $\mu_l$ and show them with empty circles in (a).}
\label{fig:fsdroplet}
\end{figure}

In Fig.~\ref{fig:fsdroplet}(a), we show the resulting chemical potential $\mu_l(\phi)$ for two different system sizes $N=4\times10^4$ and $N=2\times10^5$ setting $\gamma_{\infty}\simeq0.56$~\cite{schmitz2015ensemble}. For the smaller system we find that the droplet state is always metastable with respect to the cylinder, and only for hundred thousands of particles a clear stable plateau emerges. In simulations, the method to compute $\mu_l$ as a function of the global packing fractions $\phi$ is similar to the seeding method except that the droplet is stable ($\partial^2\Delta F/\partial V^2>0$). Therefore, after a short transient equilibrium time the evolution of the average packing fraction of the liquid phase $\phi_l$ becomes constant during a long induction time. Then, the system can either stay in the droplet state, melt, or finally move to another, more stable, geometry such as the cylinder or slab.

Determining the center of mass of the solid droplet using our bond-order parameter, we are able to measure $\phi_l$ as proposed in Refs.~\cite{statt2015finite,statt2015crystal}. In Fig.~\ref{fig:fsdroplet}(b), we show for $N=4\times10^4$ and $\phi\simeq0.51$ the typical time evolution of the packing fraction of the liquid phase $\phi_l(t)$ when the initial droplet moves to the more stable cylinder geometry. We can clearly show from $\phi_l(t)$ that it is possible to evaluate the liquid density of both the metastable droplet and the equilibrium cylinder. Since all these transitions are stochastic in nature, we perform $10$ independent runs of length $10^8\Delta t$ to average the probability distribution $P(\phi_l)$. The local maxima of this distribution are then extracted to construct the metastable and stable branches. In Fig.~\ref{fig:fsdroplet}(c), we present such a distribution and we indicate after converting $\phi_l$ into $\mu_l(\phi_l)$ the resulting values in Fig.~\ref{fig:fsdroplet}(a) (empty circles). We observe a good match of our simulations with the capillary approximation, with some deviations that we will discuss in details in the next Sec.~\ref{sec:result}.

\section{Results}
\label{sec:result}
\subsection{Finite-size droplets}

\begin{figure}[b!]
  \includegraphics[scale=1.0]{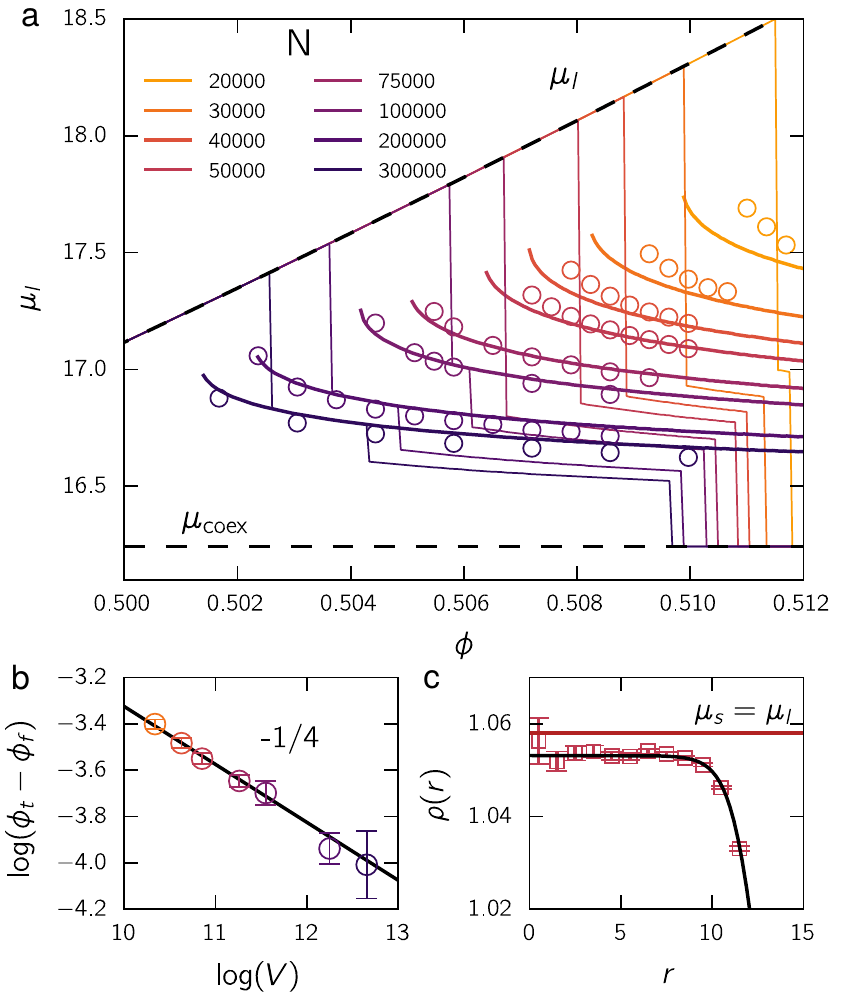}
  \caption{\textbf{Simulating finite-size droplets.} (a)~Liquid chemical potential $\mu_l$ (using the equation of state) as a function of the global packing fraction $\phi$. Empty symbols are simulation data. Thin and thick solid lines are the equilibrium and metastable droplet branch theoretical prediction (cf. Sec.~\ref{sec:fsd} for details), respectively. (b)~Logarithm of $\phi_t-\phi_f$ as a function of the logarithm of the total volume $V$. The solid line indicates the theoretical scaling with slope $-\tfrac{1}{4}$. (c)~Average radial density profile $\rho(r)$ for a solid droplet of $n\sim1\times10^4$ particles. The red solid line indicates the density of the bulk solid phase at the same chemical potential as the surrounding liquid. The black line is a fit with Eq.~(\ref{eq:rho}).}
  \label{fig:fsd}
\end{figure}

We first discuss our results concerning finite-size droplets. We vary the number of particles in the system from $N=2\times10^4$ to $N=3\times10^5$ and apply the procedure describe above. Only a single run was performed for the two largest systems ($N=2\times10^5$ and $N=3\times10^5$). In Fig.~\ref{fig:fsd}(a), we plot the chemical potential $\mu_l$ as a function of the global packing fraction $\phi$ for the various system sizes together with the prediction from minimizing the free energy Eq.~(\ref{eq:diff}). Overall, we find a good agreement between the simulations and the theoretical prediction assuming the capillary approximation. Note that it was not possible to probe metastable droplets for smaller system sizes due to the intrinsically small interfacial tension of hard spheres ($\gamma_{\infty}\simeq0.56$), which makes the interface rough and fluctuations strong. We observe a systematic deviation of the measured chemical potential (it is too big for small droplets and too small for the largest droplets), which indicates that fluctuations should be taken into account.

We find that the packing fraction $\phi_t$ at which we observe the transition to (metastable) droplets is consistent with the scaling $\phi_t-\phi_f\sim V^{-1/4}$ as shown in Fig.~\ref{fig:fsd}(b). Even more interestingly, we find for the full range of observed droplets that the density inside the droplet never reaches the bulk density $\rho_s$ at equal chemical potential of the surrounding fluid $\mu_s=\mu_l$. In Fig.~\ref{fig:fsd}(c), we show the typical density profile $\rho(r)$ and indicate the bulk density $\rho_s$ by a horizontal line. While the droplet is much larger than shown in Fig.~\ref{fig:nucleus} with a well-defined plateau inside the droplet, the density is still smaller than the bulk density.

\subsection{Coexistence}

So far, we have used the equations of state for liquid and solid branch discussed in part~\RNum{1}. Having gained access to density profiles of large droplets, we briefly return to the WCA liquid without mapping to an effective diameter. Extrapolation towards a flat interface yields estimates for the coexisting bulk densities (cf. Fig.~\ref{fig:coex}). We find that these estimates are slightly lower than previous estimates~\cite{dijkstra11}. In the following, we employ the new values yielding the chemical potential $\mu_\text{coex}\simeq16.19$ at coexistence compared to $16.24$ employed before. The effective diameter is unchanged. For more details, see appendix~\ref{sec:coex}.

\subsection{Density profiles, fluctuations, and shape}

\subsubsection{Critical density profiles}

\begin{figure*}[t]
  \includegraphics[scale=1.0]{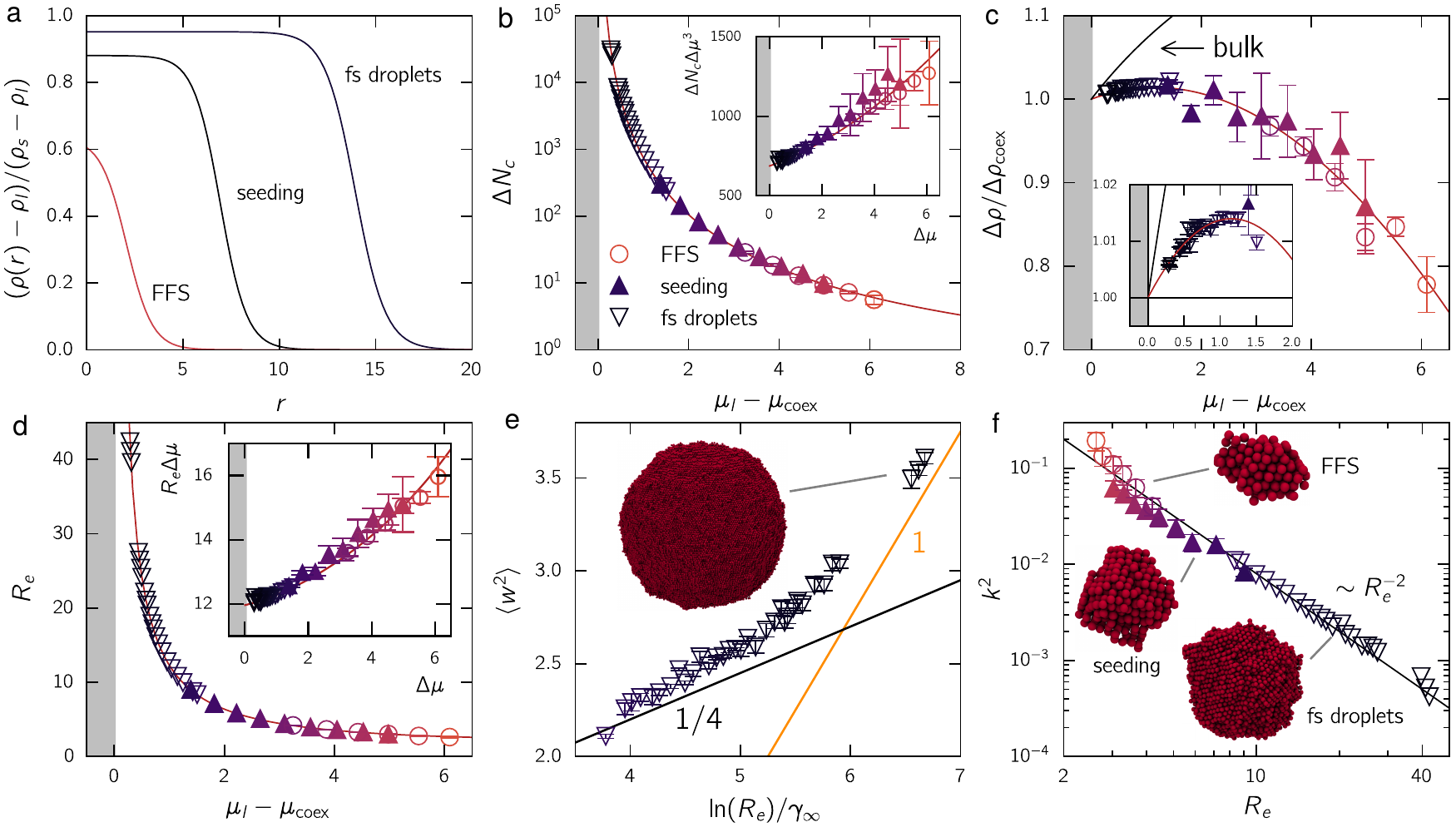}
  \caption{\textbf{Critical density profiles.} (a)~Normalized density profiles $(\rho(r)-\rho_l)/(\rho_s-\rho_l)$ of critical droplets for the three different methods: FFS, seeding, and finite-size droplets (from small to large droplets). (b) Critical excess number of particles $\Delta N_c$ as a function of the supersaturation $\Delta\mu=\mu_l-\mu_{\text{coex}}$. The solid line is a non-linear linear regression. The inset shows $\Delta N_c(\Delta\mu)^3$. (c)~Normalized density difference $\Delta\rho/\Delta\rho_{\text{coex}}$ as a function of the supersaturation. The red solid line is a non-linear regression. The black solid line shows the bulk prediction. The inset shows a zoom of the data close to the freezing point. (d)~Equimolar radius $R_e$ as a function of the supersaturation. The solid line corresponds to a combination of the fit of $\Delta N_c$ and $\Delta\rho$ through Eq.~(\ref{eq:ratio}). The inset shows $R_e\Delta\mu$. (e)~The (average) squared interfacial width $\langle w^2 \rangle$ as a function of the logarithm of the equimolar radius divided by the bulk interfacial tension $\ln(R_e)/\gamma_{\infty}$. The orange and black lines indicate the scaling $w^2\sim\frac{1}{\gamma_{\infty}}\ln R_e$ with slope 1 and 1/4, respectively. The snapshot shows the largest droplet sampled in a system composed of one million particles. (f)~The second invariant shape descriptor $k^2$ computed from the bond-order parameter as a function of $R_e$. The straight line indicates the scaling $k^2\sim R_e^{-2}$.}
  \label{fig:profiles}
\end{figure*}

We now turn to the analysis of the critical density fluctuations as a function of the supersaturation $\mu_l-\mu_{\text{coex}}$. To this end, we collect from FFS, the seeding method, and from finite size droplets the critical density profile $\rho(r)$. In Fig.~\ref{fig:profiles}(a), we show for the three different methods the critical, normalized profiles $(\rho(r)-\rho_l)/(\rho_s-\rho_l)$, where $\rho_s$ is the bulk solid density at equal ambient chemical potential $\mu_l$. We notice that, as the supersaturation decreases, the radius of the droplet increases and that the density inside the droplet $\rho(0)$ moves progressively towards its bulk value $\rho_s$ but does not reach it even for the largest droplets.

In Fig.~\ref{fig:profiles}(b), we plot the critical excess number of particles $\Delta N_c$ as a function of the degree of supersaturation $\Delta\mu=\mu_l-\mu_{\text{coex}}$. As expected, we observe a divergence of $\Delta N_c$ approaching the freezing point where $\Delta\mu\to 0$. Integrating $\partial P_i/\partial\mu_i=\rho_i$ close to coexistence for the solid and liquid branch, the pressure difference grows as $\Delta P\approx\Delta\rho_\text{coex}\Delta\mu$. Plugging this pressure difference into the nucleation work Eq.~(\ref{eq:w:cnt}) and taking the derivative with respect to $\Delta\mu$ yields [cf. Eq.~(\ref{eq:theorem})]
\begin{equation}
  \Delta N_c \approx 
  \frac{32\pi\gamma_\infty^3}{3(\Delta\rho_\text{coex})^2(\Delta\mu)^3}.
\end{equation}
Including higher orders, we fit the data against the function
\begin{equation}
  \label{eq:Nc:fit}
  \Delta N_c(\Delta\mu)^3 \approx a_n[1+b_n\Delta\mu+c_n(\Delta\mu)^2]
\end{equation}
with free coefficients $a_n$, $b_n$, and $c_n$, which is plotted in the inset of Fig.~\ref{fig:profiles}(b). We find a quasi-linear behavior, which indicates that the coefficient $c_n$ plays only a minor role. From the fit we obtain $a_n\simeq687$, which corresponds to the interfacial tension $\gamma_{n,\infty}=[3a_n(\Delta\rho_\text{coex})^2/(32\pi)]^{1/3}=0.574(7)$ for a flat interface. This value is in good agreement with the previous estimate $\gamma_{\infty}\simeq0.56$. In Fig.~\ref{fig:profiles}(b) we also show that the behavior of $\Delta N_c$ is well captured by the expression Eq.~(\ref{eq:Nc:fit}) over the full range of supersaturations studied. Moreover, the results for all three methods overlap and segue.

In Fig.~\ref{fig:profiles}(c), we show the change of the density difference $\Delta\rho=\rho(0)-\rho_l$ normalized by the difference $\Delta\rho_{\text{coex}}=\rho_s-\rho_l$ of bulk densities as a function of $\Delta\mu$. We observe a non-monotonic behavior, where for small supersaturation $\Delta\rho/\Delta\rho_{\text{coex}}$ increases slightly and then decreases. We now model the behavior by
\begin{equation}
  \frac{\Delta\rho}{\Delta\rho_{\text{coex}}}
  \approx 1 + a_{\rho}\Delta\mu + b_{\rho}(\Delta\mu)^2.
  \label{eq:drho}
\end{equation}
We find, even close to the coexistence, that the data do not follow the prediction from the bulk equation of state. Note that the pressure difference $\Delta P$ present in the nucleation work can directly be computed from the integral of $\Delta\rho(\mu)$ via Eq.~(\ref{eq:pressure}) and Eq.~(\ref{eq:ratio}). Hence, Fig.~\ref{fig:profiles}(c) implies that $\Delta P<\Delta P_{\infty}$ and thus the nucleation work from Eq.~(\ref{eq:w:cnt}) will increase even for a fixed interfacial tension due to the decrease of $\Delta P$.

The behavior of the equimolar radius $R_e$ as a function of the supersaturation is plotted in Fig.~\ref{fig:profiles}(d). As does $\Delta N_c$, this quantity diverges for $\Delta\mu\to 0$. Close the coexistence we assume the Laplace equation $\Delta P=\Delta\rho_\text{coex}\Delta\mu=2\gamma_\infty/R_e$, which implies $R_e=a_e/\Delta\mu$. This scaling is confirmed in the inset of Fig.~\ref{fig:profiles}(d). We perform a fit of $R_e$ in the same way as for $\Delta N_c$ employing a polynomial fit function, which yields $a_e\simeq12$. This coefficient should correspond to $a_e=2\gamma_{e,\infty}/\Delta\rho_\text{coex}$, from which we obtain $\gamma_{e,\infty}=0.574(1)$ again in good agreement with $\gamma_{\infty}\simeq0.56$. To model the data we employ Eq.~(\ref{eq:ratio}) combining the fits of $\Delta N_c$ and $\Delta\rho$, which is the line shown in Fig.~\ref{fig:profiles}(d).

\subsubsection{Capillary waves}

So far, we have discussed the properties of the average radial density profile $\rho(r)$, but did not discuss fluctuations present between the two coexisting phases. These fluctuations contain a bulk contribution, where both coexisting densities can fluctuate, and an interfacial contribution due do capillary waves. It is not clear for small droplets how to decouple these two effects~\cite{zhukhovitskii2008equilibrium,hofling2015enhanced}. However, one can try for large enough droplets to test if fluctuations at the surface are consistent with predictions from capillary wave theory (CWT). To this end, we have computed the square of the interfacial width $w^2$ as a function of the equimolar radius $R_e$. From CWT, we expect for a flat interface
\begin{equation}
  w^2 \sim \frac{1}{4\gamma_{\infty}}\ln L
  \label{eq:cap}
\end{equation}
as a function of the length $L$ of the interface~\cite{vink2005capillary,zykova2010monte}. This relation has been used to extract the interfacial tension of a Lennard-Jones fluid close to criticality~\cite{watanabe2012phase}. For spherical droplets less numerical results are available. Note that Eq.~(\ref{eq:cap}) was previously used to analyze spherical droplets in a lattice gas model through replacing $L$ by the circumference of the droplet~\cite{schmitz2013monte}. In this study, the authors found that the (inverse) slope $4\gamma_{\infty}$ is slightly lower compared to previous calculations of the interfacial tension. Two theoretical studies have derived a possible scaling of $w^2$~\cite{pavloff1998rough,prestipino2014shape}. In Ref.~\cite{prestipino2014shape}, the authors derived $w^2\sim\frac{1}{3\gamma_{\infty}}\ln V_s$, which gives
\begin{equation}
  w^2\sim\frac{1}{\gamma_{\infty}}\ln R_e.
\end{equation}
In Fig.~\ref{fig:profiles}(e), we plot $w^2$ against $\ln(R_e)/\gamma_{\infty}$ to test these possible scalings. For smaller droplets ($R_e<15$), we find that our data is consistent with slope $1/4$ but for larger droplets the slope increases and seems to reach a slope of unity ($R_e>20$). However, one would need even larger droplets to confirm this scaling.

\subsubsection{Droplet shape}
\label{sec:shape}

Since we have adopted a spherical symmetry to evaluate the density profiles, it is also of interest to quantify to which extend this spherical assumption is valid. The deviation of sampled droplets from a spherical shape can be quantified via either the bond-order parameter or by constructing an iso-density surface. We choose to compute, for $n$ particles identified as solid and ``bonded'' to from the droplet, the gyration tensor~\cite{arkin2013gyration}
\begin{equation}
  \label{eq:Rg}
  \mathbf R_g^2 = \frac{1}{n}\sum_{i=1}^{n}(\mathbf r_i
  -\bar{\mathbf r})\otimes(\mathbf r_i-\bar{\mathbf r})
\end{equation}
with center-of-mass $\bar{\mathbf r}$ of the droplet. From its eigenvalues $\lambda_i$ one calculates the second invariant shape descriptor $k^2=1-3(\lambda_1 \lambda_2 + \lambda_2 \lambda_3 + \lambda_1 \lambda_3)/(\lambda_1+\lambda_2+\lambda_3)^2$, which quantifies how the droplet shape deviates from a sphere with $k^2=0$. The upper bound of $k^2$ corresponds to a rod with $k^2=1$. In Fig.~\ref{fig:profiles}(f), we present the average $\mean{k^2}$ as a function of the equimolar radius $R_e$ of the droplet. For finite-size droplets, we find that $\mean{k^2}$ decreases with $R_e$ and scales as $\sim R_e^{-2}$ as expected for a spherical shape perturbed by fluctuations (see appendix~\ref{sec:scaling}). We observe that this scaling extends to the smaller droplets sampled by seeding and FFS simulations, which suggests that the droplets remain basically spherical although single droplets look of course more fuzzy (see the snapshots). Only for the smallest droplets do we observe a significant deviation from the scaling. The same qualitative behavior was reported for one-component and binary Lennard-Jones crystalline droplets~\cite{jungblut2011crystallization}. The small gap between seeded droplets and FFS droplets indicates that the shape of the seeded droplets (which start as perfectly spherical) has not fully relaxed.

\subsection{Nucleation work}

\begin{figure}[t]
  \includegraphics[scale=1.0]{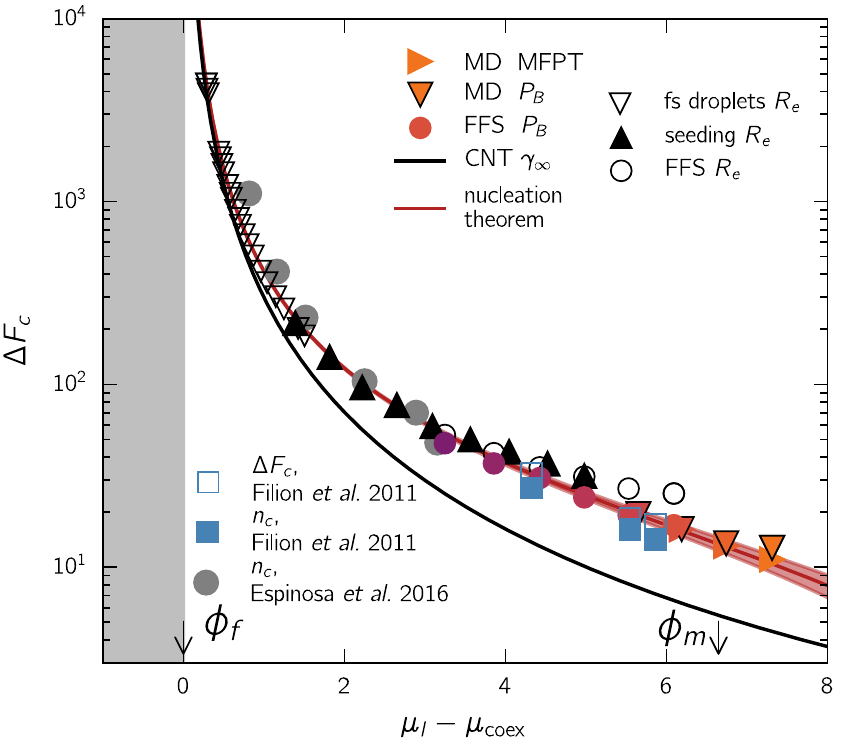}
  \caption{\textbf{Nucleation work.} Nucleation work $\Delta F_c$ as a function of the supersaturation $\mu_l-\mu_{\text{coex}}$ obtained from unbiased MD simulations ($\blacktriangledown,\blacktriangleright$), FFS ($\bullet$), and umbrella sampling ($\square$, Ref.~\cite{dijkstra11}). The black solid line is the CNT prediction $\Delta F_{c,\infty}$ based on the bulk pressure difference $\Delta P_\infty$ and $\gamma_\infty\simeq0.56$. The thick red solid line is obtained from the nucleation theorem Eq.~(\ref{eq:ntheorem}) by averaging different reference state points for $\mu_0$ from MD and FFS data. The red shaded area shows the corresponding standard deviation. For comparison, we also show the nucleation work $\Delta F_{c,e}$ calculated from the equimolar radius $R_e$ ($\triangledown$,$\blacktriangle$,$\circ$) and $\Delta F_{c,\mu}$ based on the data from Ref.~\cite{espinosa2016seeding} (gray discs) and Ref.~\cite{dijkstra11} (blue squares).}
  \label{fig:work}
\end{figure}

In Fig.~\ref{fig:work}, we present the nucleation work $\Delta F_c$ as a function of the supersaturation $\mu_l-\mu_\text{coex}$. It can be calculated from unbiased MD and FFS simulations as described in part~\RNum{1}~\cite{speckpartone}. For the smaller system sizes employed in FFS, we compute $\mu_l$ from the liquid density $\rho_l$ extracted from the critical density profile and not the global density in order to correct for the depletion of the liquid phase as the droplet forms (cf. the finite-size effect in Fig.~\ref{fig:seeding}(d) for the lowest packing fraction $\phi\simeq0.52$). We observe that the CNT prediction $\Delta F_{c,\infty}$ employing $\gamma_\infty$ and $\Delta P_\infty$ underestimates the true nucleation work.

\begin{figure*}[t]
  \includegraphics[scale=1.0]{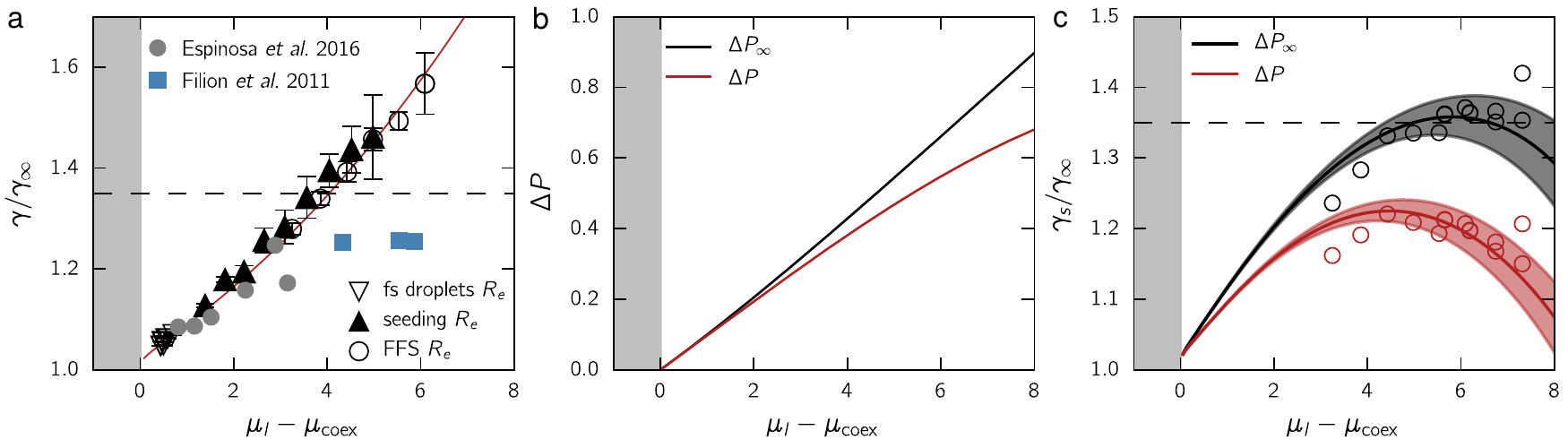}
  \caption{\textbf{Interfacial tension.} (a)~Estimation of the normalized interfacial tension $\gamma/\gamma_{\infty}$ as a function of $\mu_l-\mu_{\text{coex}}$ through inverting the CNT expression for (i) using the equimolar radius ($\gamma_e\simeq\Delta P_{\infty}R_e/2$) applied to our data, and (ii) $\gamma_\mu$ (see main text) applied to Refs.~\cite{dijkstra11,espinosa2016seeding}. (b)~Pressure difference between the solid droplet and ambient liquid as a function of $\mu_l-\mu_{\text{coex}}$. The solid and red lines are the bulk ($\Delta P_{\infty}$) and true ($\Delta P$) pressure difference, respectively. (c)~Normalized interfacial tension $\gamma_s/\gamma_{\infty}$ as a function of supersaturation $\mu_l-\mu_{\text{coex}}$. The black and red symbols are obtained by inverting Eq.~(\ref{eq:w:cnt}) using the nucleation work ($\Delta F_c$) and the bulk ($\Delta P_{\infty}$) and true ($\Delta P$) pressure difference, respectively. The thick lines are obtained from the nucleation theorem (cf. Fig.~\ref{fig:work}). Empty circles are direct estimates from the MD and FFS data. The shaded areas show the corresponding standard deviation. The dashed horizontal lines indicate $\gamma_\text{CNT}/\gamma_\infty\approx1.35$.}
  \label{fig:surface}
\end{figure*}

\begin{table}[b!]
  \centering
  \begin{tabular}{c|c|c|c|c}
    & $\gamma_s$ & $\gamma_\infty$ & $R_e$ & $n_c$ \\
    \hline\hline
    $\Delta P$ & $\Delta F_c$ & & \\
    \hline
    \multirow{2}{*}{$\Delta P_\infty$} & & $\Delta F_{c,\infty}$ & $\Delta F_{c,e}$ & $\Delta F_{c,\mu}$ \\
    & $\gamma_\text{CNT}$ & & $\gamma_e$ & $\gamma_\mu$
  \end{tabular}
  \caption{Summary of the different nucleation works based on the different input quantities (top row and left column). The last line indicates the resulting effective surface tensions.}
  \label{tab:work}
\end{table}

Having computed $\Delta F_c$ for several $\mu_l$ and evaluated $\Delta N(\mu_l)$ from various methods, we can now apply the nucleation theorem. We first find that, independent of the starting point $\mu_0$ of the integration through Eq.~(\ref{eq:ntheorem}), all computed curves $\Delta F_c(\mu_l)$ lie on top of each other. Moreover, they slowly converge towards $\Delta F_{c,\infty}$ approaching the freezing point in agreement with the fact that the curvature $1/R\to0$ vanishes. This highlights that the capillary approximation is in practice only valid for barriers $\Delta F_c>1000k_BT$. This regime corresponds to droplets composed of thousands of particles, which is beyond what rare event sampling methods can achieve.

Very often the droplet radius $R_s$ corresponding to the surface of tension is not known and the droplet radius is estimated from the cluster size $n_c$ of solid particles. Besides the issue that $n_c$ depends on the specific choice of order parameter, it neither corresponds to $R_s$ nor $R_e$ although in practice it might be close. To see the impact of varying the droplet radius, we have calculated the nucleation work $\Delta F_{c,e}=\Delta P_{\infty}V_s(R_e)/2$. For hard spheres, we find a good agreement between this approximation and the work estimated from the nucleation theorem for $\Delta F_c>50k_BT$, but for smaller barriers it overestimates the nucleation work. We also compared our data with seeding results from Ref.~\cite{espinosa2016seeding} obtained in the NPT ensemble. In this study, the authors employ $\Delta F_{c,\mu}=n_c(\mu_l-\mu_s)/2$ at fixed pressure [cf. Eq.~(\ref{eq:w:mu})] to calculate the nucleation work. As already discussed, one thus makes additional approximations through using the difference between bulk solid and liquid chemical potential and assuming that the thermodynamic variable $N_s\approx n_c$ corresponds to the order parameter. Despite these approximations, for hard spheres we find an overall good agreement between this approach and the results from the nucleation theorem, with deviations for large barriers $\Delta F_c>100k_BT$. However, that one has to be careful is demonstrated by using the data from Filion \emph{et al.}~\cite{dijkstra11}, for which the nucleation work is now underestimated. The different expressions for the nucleation work are summarized in Table~\ref{tab:work}.

\subsection{Interfacial tension and pressure difference}

Finally, we turn to the interfacial tension and pressure difference between the solid droplet and the ambient liquid. A common approach to extract the interfacial tension is to employ CNT assuming the bulk difference of pressure between the solid and liquid phase $\Delta P_{\infty}$. Additionally, the second approximation is to employ the (directly accessible) droplet size $n_c$ defined by a bond order parameter instead of the correct radius of tension $R_s$. Non-classical effects are then subsumed into an effective interfacial tension, either $\gamma_e=\Delta P_{\infty} R_e/2$ or $\gamma_\mu=[(3\rho_s^2n_c)/(32\pi)]^{1/3}(\mu_l-\mu_s)$ for the data from Refs.~\cite{dijkstra11,espinosa2016seeding}. In Fig.~\ref{fig:surface}(a), we plot the thus obtained interfacial tensions $\gamma_i/\gamma_{\infty}$ (with $i=e,\mu$) normalized by $\gamma_\infty$ as a function of the supersaturation $\Delta\mu$. We observe an approximately linear increase. For $\Delta\mu\to0$, we recover the interfacial tension $\gamma_{\infty}$ of a flat interface from the data.

For larger supersaturations and smaller droplets, the interfacial tension keeps increasing. Fitting the numerical nucleation rates, in part~\RNum{1} we found the effective interfacial tension $\gamma_\text{CNT}\approx1.35\gamma_\infty$ over a range $3.7<\Delta\mu<7.3$. Clearly, this (approximately constant) result does not agree with the effective interfacial tensions extracted from the two versions $\Delta F_{c,e}$ and $\Delta F_{c,\mu}$ of the approximated nucleation work. This demonstrates the inconsistency of these approximations and their failure to predict the nucleation rate. This failure can be traced to the pressure inside droplets diverting from the bulk prediction. This is shown in Fig.~\ref{fig:surface}(b), where we plot the difference $\Delta P_\infty$ between the bulk phases and the actual pressure difference $\Delta P$ calculated via Eq.~(\ref{eq:pressure}) from the excess number of solid particles. We find that close to the coexistence the two pressures are the same for a given ambient chemical potential $\mu_l$ but for larger supersaturation the pressure inside a finite droplet becomes smaller compared to the solid bulk phase. For a more detailed discussion on this scenario, see Ref.~\cite{kashchiev2003thermodynamically}. This highlights why the simple approximation $\gamma_e=\Delta P_{\infty} R_e/2$ leads to an incorrect estimate of the interfacial tension.

Having computed both the nucleation work $\Delta F_c$ and the pressure difference $\Delta P$, we can finally extract the surface of tension $\gamma_s$ through inverting Eq.~(\ref{eq:w:cnt}). In Fig.~\ref{fig:surface}(c), we plot the ratio $\gamma_s/\gamma_{\infty}$ as a function of the supersaturation $\Delta\mu$. For $\Delta\mu\to0$ we again recover the interfacial tension of the flat interface. The maximal tension is $20\%$ larger than $\gamma_\infty$ but decreases again for very large supersaturations where the nucleations barrier vanishes. For comparison, we also show the interfacial tension using the extrapolated nucleation work but employing the bulk pressure difference $\Delta P_\infty$. The qualitative behavior is similar but the apparent surface tension of droplets is much larger and now plateaus at $\gamma_\text{CNT}$ in agreement with the value determining the nucleation rates.

\section{Conclusion}

In this work, we have combined four numerical methods: unbiased MD, FFS, the seeding method, and finite-size droplets to study nucleation in the hard-sphere model. We found that these methods are complementary to each other and allow to study nucleation over a wide range of supersaturations. All methods sample the same ensemble of critical droplets, which can be rationalized by the time scale separation between slow nucleation kinetics and the fast structural relaxation within droplets. The reversible work $\Delta F_c=-\Delta PV_s+\gamma A$ comprises a bulk contribution due to the pressure difference $\Delta P$ between the droplet and ambient phase, and an excess (surface) free energy determined by the interfacial tension $\gamma$. In classical nucleation theory (CNT), two crucial approximations are employed: the pressure difference is set to the difference $\Delta P_{\infty}$ between bulk phases, and the interfacial tension is set to $\gamma_{\infty}$ of a flat interface. From the computed density profiles, we quantify the first non-classical effect affecting the bulk contribution: The density at the center of the droplet reaches its bulk value only in the thermodynamic limit. As a consequence, the pressure difference between solid droplet and metastable liquid is lower than the bulk pressure difference $\Delta P_{\infty}$.

From the density profiles, we also obtain the excess number of particles and the equimolar radius, which we model from thermodynamic arguments. Calculating interfacial tensions from computer simulations is notoriously difficult even for flat interfaces. This is particularly true for hard spheres, where the tension is small and purely entropic. We have shown here that extrapolating the interfacial tension of droplets to infinite size yields the value $\gamma_\infty\simeq0.57$ independent of the approximations involved. This value is in good agreement with the value reported in Ref.~\cite{schmitz2015ensemble}, see also the discussion therein.

To compute the nucleation work for the supersaturations studied, we combine the nucleation theorem with barriers extracted from unconstrained dynamics (FFS and MD). The advantage of this new method is that it does not rely on approximations. From the extracted barriers, we again conclude that the CNT prediction assuming the bulk pressure difference $\Delta P_{\infty}$ and flat interfacial tension $\gamma_{\infty}$ is only valid in the thermodynamic limit. In the seeding and finite-size droplets method, the nucleation barrier is not accessible and one relies on using the CNT expression to evaluate the work. Here, we demonstrate that this route does not yield reliable results, although it becomes more accurate for large barriers (large droplets). Beside employing the bulk pressure difference, a second source of error is approximating the droplet volume associated with the radius of tension with an ``arbitrary'' volume, \emph{e.g.}, from a bond order parameter or the lever rule based on the bulk solid equation of state.

To resolve these issues, we make use of a variation of the original nucleation theorem by Kashchiev~\cite{kashchiev1982relation,kashchiev2000nucleation}, which links the variation of the pressure difference as a function of the ambient chemical potential to the density inside the solid droplet. This thermodynamic route bypasses the difficulties to employ the mechanical route from the pressure tensor, which is ill-defined for small droplets~\cite{schofield1982statistical,varnik2000molecular}. We are then able to quantity a second non-classical effect, namely the increase of the excess free energy for finite droplets. We find that both a decrease of the pressure difference and an increase of the interfacial tension are responsible for the increase of the nucleation work compared to the bulk approximation of CNT.

The method presented in this paper provides an elegant way to both extract nucleation barriers and to disentangle non-classical effects, \emph{i.e.}, to quantify the deviations from the bulk equations of state in inhomogeneous systems with curved interfaces. We have studied a simple model system, hard spheres, but our insights regarding the use of CNT are general and apply to a large class of phase transformation kinetics including the nucleation of ice~\cite{li2011homogeneous,sanz2013homogeneous,russo2014new,haji2015direct,espinosa2016interfacial,sosso2016crystal,lupi2017role}.


\begin{acknowledgments}
  We thank K. Binder for helpful discussions and suggestions. We gratefully acknowledge ZDV Mainz for computing time on the MOGON supercomputer. We acknowledge financial support by the DFG through the collaborative research center TRR 146.
\end{acknowledgments}

\appendix

\section{Bond-order parameter}
\label{ap:bop}

The local bond-orientational parameter is computed as in part~\RNum{1} and closely follows Ref.~\cite{filion2010crystal}. First, the expression 
\begin{equation}
  \label{eq:order}
  q_{l,m}(i) = \frac{1}{N_n(i)}\sum_{j=1}^{N_n(i)} Y_{l,m}(\theta_{i,j},\varphi_{i,j})
\end{equation}
is evaluated for particle $i$, where $Y_{l,m}(\theta,\varphi)$ are spherical harmonics and $N_n$ is the number of neighbors within distance $r_{ij} < 1.5\sigma$. We then construct a bond network through the scalar product
\begin{equation}
  \label{eq:scalar}
  d(i,j) = \frac{\sum_{m=-l}^{l} q_{l,m}(i) q_{l,m}^\ast(j) }{(\sum_{m=-l}^{l} |q_{l,m}(i)|^2)^{1/2} (\sum_{m=-l}^{l} |q_{l,m}(j)|^2)^{1/2}}
\end{equation}
using $l=6$ with $d(i,j) > 0.7$ defining a bond. A particle is defined as "solid-like" if the number of bonds $\xi\ge 9$, and clusters are constructed from mutually bonded solid-like particles.

\section{Constructing density profiles}
\label{ap:profile}

\begin{figure}[t]
  \includegraphics[scale=1.0]{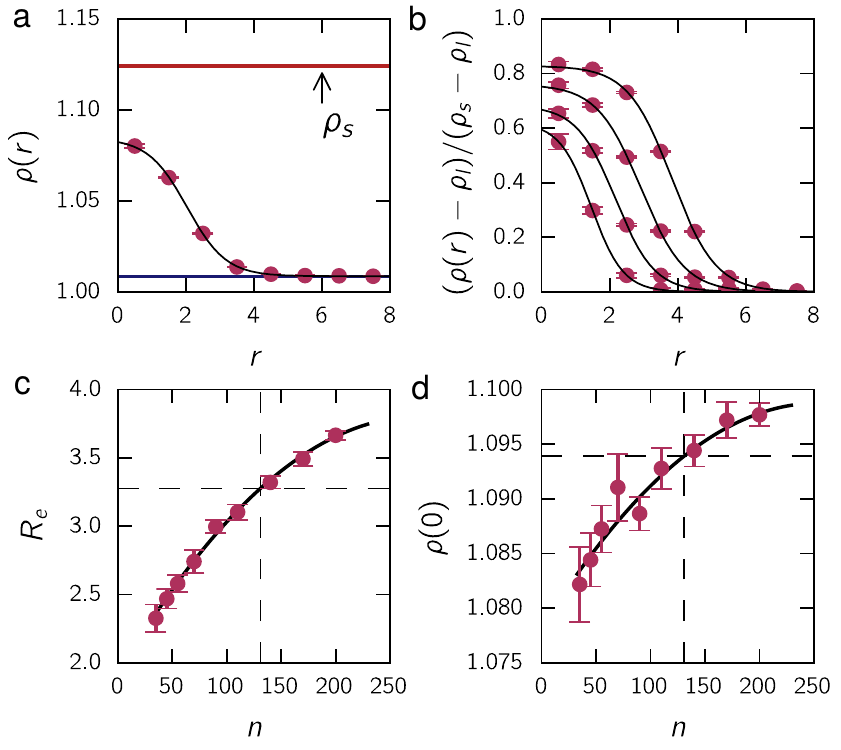}
  \caption{\textbf{Density profiles construction.} (a)~Radial density profile $\rho(r)$ averaged in FFS over configurations at an interface placed at $n=55$ and $\phi\simeq0.529$. (b)~Evolution of the normalized density profile $(\rho(r)-\rho_l)/(\rho_s-\rho_l)$ for various interfaces. All solid lines in (a) and (b) corresponds to a fit from Eq.~(\ref{eq:rho}). (c)~Equimolar radius $R_e$ as a function of the droplet size $n$. (d)~Solid density in the center of the droplet as a function of the droplet size $n$. Solid lines in (c) and (d) are polynomial fit to extract the critical value indicated by dashed lined, where $n_c$ is determined via the committor probability.}
  \label{fig:rho}
\end{figure}

Here, we describe the method used to extract the critical droplet profiles in FFS and in the seeding method. For every configurations at a given $n$ computed from our bond order parameter, we compute the voronoi volume $v_i$ of each particle $i$. We then compute the center of mass of the solid droplet and evaluate the density profile $\rho(r)$ by binning with $\Delta r=d_\text{eff}$ the quantity $\rho(r_i)=1/\langle v_i\rangle$, where the $v_i$ are taken from particles inside the bin centered at $r_i$. In Fig.~\ref{fig:rho}(a), we show such a profile for configurations generated from FFS at $\phi\simeq0.529$ and $n=55$. The profile is well fitted by the mean-field expression
\begin{equation}
  \rho(r)=\frac{\rho_l+\rho(0)}{2}+\frac{\rho_l-\rho(0)}{2}
  \tanh \left(\frac{r-R_0}{w}\right),
  \label{eq:rho}
\end{equation}
where $\rho(0)$ is the density at the center of the solid droplet, $\rho_l$ is the density of the surrounding liquid, $R_0$ is the radius at half maximum, and $w$ the interfacial width. The equimolar radius $R_e$ is calculated from the absorption
\begin{equation}
  \label{eq:eds}
  R^2\Gamma_x(R) =\int_{0}^{R}dr\;r^2[\rho(r)-\rho(0)]
  - \int_{R}^{\infty}dr\;r^2[\rho(r)-\rho_l]
\end{equation}
through the condition $\Gamma_x(R_e)=0$.

In Fig.~\ref{fig:rho}(b), we show the dependence of the normalized density profile $(\rho(r)-\rho_l)/(\rho_s-\rho_l)$ for various $n$ corresponding to different interfaces in FFS. We observe during the crystal growth an increase of the droplet radius in addition to a densification of its center. We finally can extract information about the critical density fluctuation by fitting the quantity of interest, let say the equimolar radius, as function of the droplet size. In practice, we use a polynomial function to model an observable $A(n)$ between $0.25n_c<n<0.75n_c$ and then extract $A(n_c)$ from the resulting fit. We show two examples of this procedure in Fig.~\ref{fig:rho}(c) and (d) for the equimolar radius $R_e$ and the solid density at the center of the droplet $\rho(0)$, respectively. For equilibrium droplets present at the droplet transition, we evaluate the profile without any interpolations and simply average $\rho(r)$ over the metastable runs far away from any transitions to a melt or a cylinder.

\section{Modified equations of state}
\label{sec:coex}

\begin{figure}[b!]
  \includegraphics[scale=1.0]{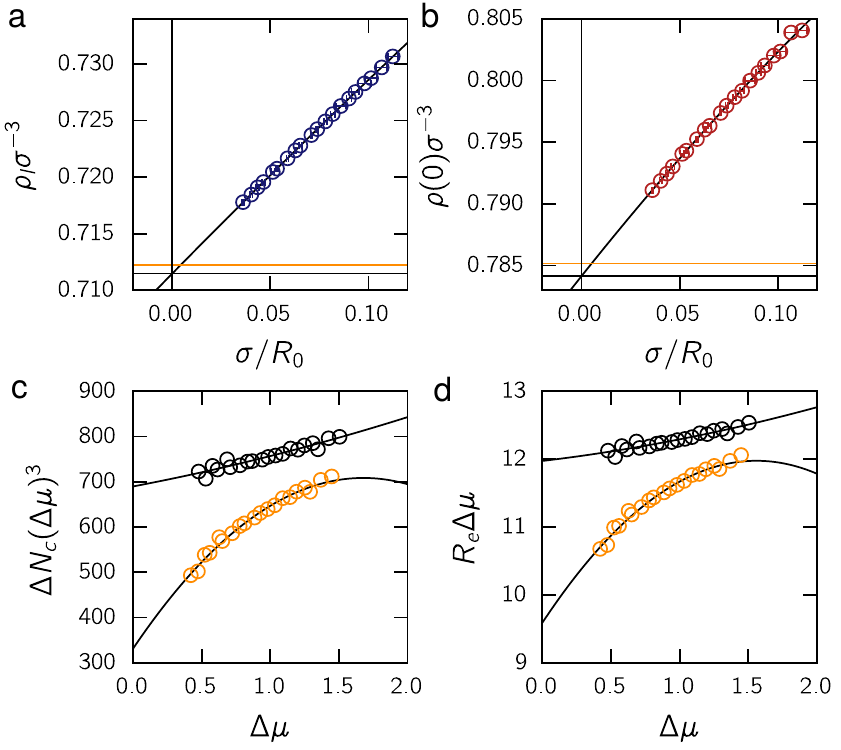}
  \caption{\textbf{Finite-size droplets.} (a)~Liquid density $\rho_l$ and (b)~solid density $\rho(0)$ of finite-size droplets (in WCA units) as a function of the droplet curvature $\sigma/R_0$. The solid lines are polynomial fits of second order. The black and orange horizontal lines are the bulk densities extracted from the fits and the bulk prediction from the equations of state, respectively. (c)~Scaled excess number $\Delta N_c(\Delta\mu)^3$ and (d)~scaled equimolar radius $R_e\Delta\mu$ as a function of $\Delta\mu=\mu_l-\mu_{\text{coex}}$ (in hard sphere units). Orange and black data are from the old and new freezing point, respectively.}
  \label{fig:coex}
\end{figure}

From fitting the density profiles, we obtain the liquid density $\rho_l$ and the droplet density $\rho(0)$, which are plotted in Fig.~\ref{fig:rho} as a function of the droplet curvature $1/R_0$. Extrapolating these densities towards a flat interface $1/R_0\to0$ should yield the densities of bulk liquid and solid at coexistence. These densities have been determined previously in Ref.~\cite{dijkstra11} from simulations with $N=4000$ particles. For the droplets we find slightly lower values, $\rho_f\simeq0.71147(4)$ and $\rho_m\simeq0.7841(1)$. Converting to pressure using the equations of state, this would imply a pressure difference $\beta\sigma^3\Delta P\simeq-0.0003$. To avoid this inconsistency, we have decided to re-parametrize the solid equation of state by shifting $P_s(\rho)$ so that $P_s(\rho_m)=P_l(\rho_f)$. This yields a coexistence pressure $P_\text{coex}=11.639(3)$ in good agreement with previous estimates~\cite{fernandez2012,espinosa2013}. We have then performed a thermodynamic integration to evaluate $\mu_s$, taking the coexistence as new reference state point where $\mu_s(\rho_m)=\mu_l(\rho_f)$. Old and updated values are summarized in Table~\ref{tab:coex}.

\begin{table}[t]
  \centering
  \begin{tabular}{c|c|c|c|c}
    & $\rho_f\sigma^{-3}$ & $\rho_m\sigma^{-3}$ & $P_{\text{coex}}$ & $\mu_{\text{coex}}$ \\
    \hline\hline
    previous & 0.712 & 0.785 & 11.70 & 16.24 \\
    \hline
    new & 0.71147(4) & 0.7841(1) & 11.639(3) & 16.185(3)
  \end{tabular}
  \caption{Freezing density $\rho_f$, melting density $\rho_m$, pressure, and chemical potential at coexistence. The first row defines the parametrization of the bulk equations of state using the values provided in Ref.~\cite{dijkstra11}. The bottom line shows the updated values based on the extrapolation of droplets to vanishing curvature.}
  \label{tab:coex}
\end{table}

Fig.~\ref{fig:coex}(c,d) show the extrapolation of the excess number $\Delta N_c$ and equimolar radius for $\Delta\mu\to0$ [cf. Fig.~\ref{fig:profiles}(b,d)]. Keeping the old coexisting densities, these extrapolations would imply interfacial tensions $\gamma_{n,\infty}\simeq0.451(5)$ and $\gamma_{e,\infty}\simeq0.461(3)$, respectively. Both values disagree with each other (within errors) and, more importantly, are to small compared to independent estimations of the interfacial tension of flat interfaces. In contrast, for the updated freezing point we obtain $\gamma_{n,\infty}\simeq0.574(5)$ and $\gamma_{e,\infty}\simeq0.574(3)$, which are now in excellent agreement with each other and previous estimates.

\section{Scaling of shape descriptor}
\label{sec:scaling}

We give a simple argument how the shape descriptor used in Sec.~\ref{sec:shape} is influenced by thermal fluctuations. The thermal expectation reads
\begin{equation}
  \mean{k^2} \propto \int[\dd u]\; k^2[u] e^{-\beta\delta F[u]}
\end{equation}
up to normalization. The path integral sums over all fluctuations of the droplet's interface with the liquid, the position of which is described by $\X(\vhi,\theta)=R(1+u)\mathbf e_r$ with a small perturbation $u(\vhi,\theta)$ of a sphere with radius $R$. Considering droplets with a homogeneous density, the gyration tensor Eq.~(\ref{eq:Rg}) can be written
\begin{equation}
  \label{eq:Rg:int}
  \Int{^3\x}\frac{\mathbf e_r\otimes\mathbf e_r}{V_s}
  = \IInt{\vhi}{0}{2\pi}\IInt{\theta}{-\pi}{\pi}
  \sin\theta\frac{(1+u)^3}{4\pi}(\mathbf e_r\otimes\mathbf e_r).
\end{equation}
The free energy of shape fluctuations away from the average spherical shape (and, in principle, at constant volume $V_s$) generically reads $\delta F[u]\approx\tfrac{1}{2}\alpha(A[u]-A_0)^2$ with coefficient $\alpha>0$ and $A_0=4\pi R^2$. The area to lowest order is
\begin{equation}
  A[u] \approx \IInt{\vhi}{0}{2\pi}\IInt{\theta}{-\pi}{\pi} 
  R^2\sin\theta(1+2u).
\end{equation}
The shape descriptor is build from the eigenvalues of instantaneous gyration tensors and then averaged. The average can be expressed as a Gaussian path integral over all fluctuations $u$. Expanding the shape descriptor, the lowest-order contribution is $\mean{k^2}\sim\mean{u^2}$ and thus
\begin{equation}
  \mean{k^2} \sim \frac{1}{\sqrt{\beta\alpha R^4}} \sim R^{-2}.
\end{equation}
This shows that the scaling in Fig.~\ref{fig:profiles}(f) can be attributed to fluctuations and does not indicate a departure from the basically spherical shape of droplets.


\end{document}